\documentclass[3p,times,twocolumn]{elsarticle}
 \biboptions{comma,sort&compress}
 
\usepackage{graphicx}
\usepackage{amsmath,amssymb}
\usepackage{here}
\usepackage{ecrc}

\volume{00}

\firstpage{1}

\journalname{Nuclear and Particle Physics Proceedings}

\runauth{}

\jid{nppp}

\jnltitlelogo{Nuclear and Particle Physics Proceedings}

\usepackage{amssymb}
\usepackage[figuresright]{rotating}

\begin{document}

\begin{frontmatter}

%%
%%%%%%%%%%%%%%%%%%%%%%%%%%%%%%%%%%%%%%%%%%%%%%%%%
\title{ On the high-energy instability of quarkonium production$^*$}
 % \corref{cor0}}
 \cortext[cor0]{Talk given at 26th International Conference in Quantum Chromodynamics (QCD23),  10-14 july 2023, Montpellier - FR}
 \author[label1]{M.A. Nefedov}
%  \cortext[cor0]{FAPESP CNPq-Brasil PhD student fellow.}
\ead{maxim.nefedov@ijclab.in2p3.fr}
\address[label1]{Universit\'e Paris-Saclay, CNRS, IJCLab, 91405 Orsay, France}

\pagestyle{myheadings}
\markright{ }
\begin{abstract}
The perturbative instability of NLO collinear factorisation (CF) computations of $p_T$-integrated cross sections of heavy quarkonium production at high hadronic or photon-hadron collision energy is discussed. We resolve this problem via the matching of NLO CF computation with the resummation of higher-order corrections $\propto \alpha_s^n \ln^{n-1}(\hat{s}/M^2)$ at high partonic center of mass energies $\hat{s}\gg M^2$. The resummation is performed in the Doubly-Logarithmic Approximation(DLA) of High-Energy Factorisation(HEF) formalism. We also report the results of the first computation of one-loop corrections to impact-factors involving heavy quark-antiquark pair in the intermediate states considered in the Non-Relativistic QCD (NRQCD) factorisation formalism for quarkonium production: $Q\bar{Q}\left[{}^1S_0^{[8]} \right]$ and $Q\bar{Q}\left[{}^1S_0^{[1]} \right]$. These results are necessary for the extension of our resummation formalism beyond DLA.
\end{abstract}
% \begin{document}

\end{frontmatter}
%%%%%%%%%%%%
%\vspace*{-1.5cm}

\section{High-Energy instability of quarkonium production cross sections and HEF${}^1$}
\footnotetext[1]{This part is based on the work done in collaboration with Jean-Philppe Lansberg and Melih Ozcelik~\cite{Lansberg:2021vie, Lansberg:2023kzf}.}

\begin{figure}[t]
\begin{center}\includegraphics[width=0.4\textwidth]{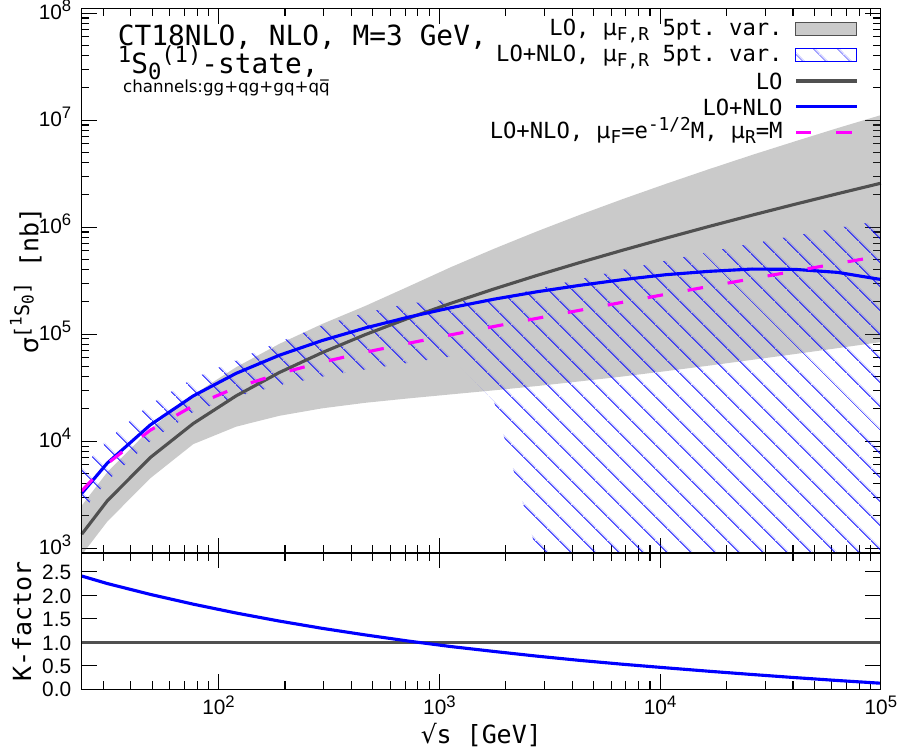}\end{center}
\caption{The $pp$ collision energy($\sqrt{s}$) dependence of the total cross section of production of the $c\bar{c}$-pair in the ${}^1S_0^{[1]}$ state in the LO (gray curve) and NLO (blue curve) of CF, shown together with the corresponding 5-point scale-variation bands. The NLO computation with the $\hat{\mu}_F$-scale of Ref.~ is shown by the dashed line. The figure is taken from Ref.~. }\label{fig:sig-etac-CF}
\end{figure}

\begin{figure}[t]
\begin{center}\includegraphics[width=0.4\textwidth]{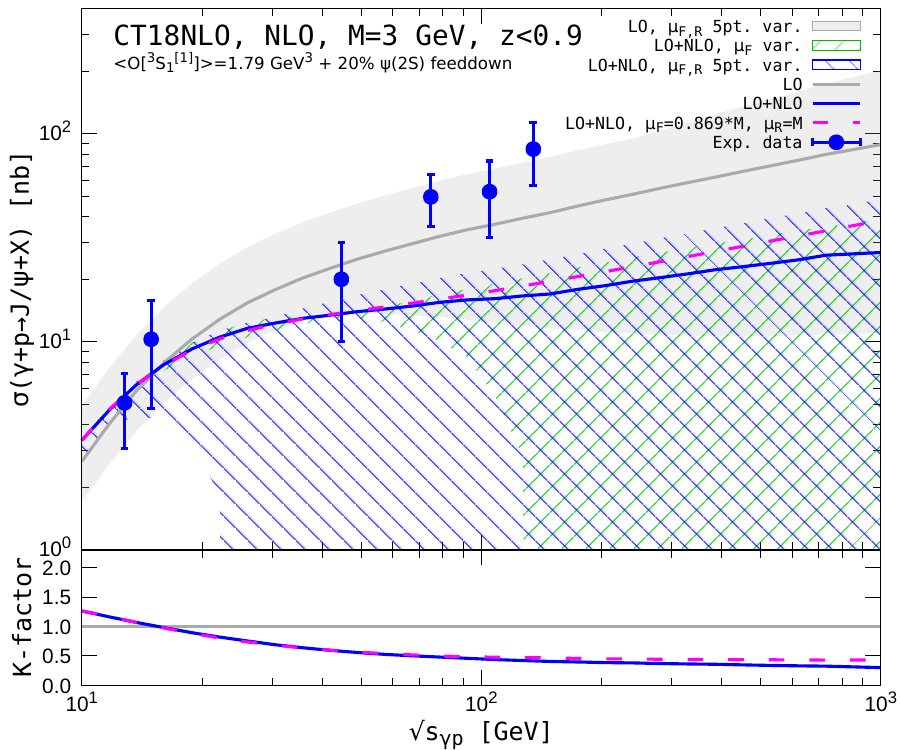}\end{center}
\caption{The $\gamma p$ collision energy($\sqrt{s_{\gamma p}}$) dependence of total cross section of prompt $J/\psi$ photoproduction in the CSM at LO (grey curve) and NLO (blue curve) of CF, shown together with the corresponding 5-point scale-variation bands. The NLO computation with the $\hat{\mu}_F$-scale of Ref.~ is shown by the dashed line. The figure is taken from Ref.~. }\label{fig:sig-jpsi-CF}
\end{figure}

 Since the heavy quarkonium mass $M$($\simeq 2m_c$ or $2m_b$) provides a hard scale, the computation of $p_T$-integrated quarkonium production cross sections should in principle be possible using perturbative QCD combined with standard collinear factorisation (CF) theorem for initial state as well as Non-Relativistic QCD (NRQCD) factorisation hypothesis~\cite{Bodwin:1994jh} to describe the hadronisation of heavy quark-antiquark pair ($Q\bar{Q}$) into quarkonium. However, as it was emphasized in recent papers~\cite{Lansberg:2020ejc,ColpaniSerri:2021bla}, such CF computation develops an extremely strong sensitivity to the choice of factorisation scale $\mu_F$, when hadronic or photon-hadron collision energy becomes large in comparison to $M$. The plot in the Fig.~\ref{fig:sig-etac-CF} illustrates this phenomenon for the case of inclusive $\eta_c$ hadroproduction cross section, studied in Ref.~\cite{Lansberg:2020ejc} in the approximation that the $\eta_c$ production is dominated by the $c\bar{c}\left[{}^1S_0^{[1]} \right]$ NRQCD intermediate state. As one can see, the usual scale-variation band of the NLO computation explodes for $\sqrt{s}> 1$ TeV and one can get negative cross sections at high $pp$ collision energy for reasonable choice of scales. In the Fig.~\ref{fig:sig-jpsi-CF} we observe the similar behaviour of inclusive $J/\psi$ photoproduction cross section for $\sqrt{s_{\gamma p}}>20$ GeV, which was studied in Ref.~\cite{ColpaniSerri:2021bla} in the CS approximation of dominating $c\bar{c}\left[{}^3S_1^{[1]}\right]$ state. Lifting the colour-singlet (CS) approximation of Refs.~\cite{Lansberg:2020ejc,ColpaniSerri:2021bla} will not resolve this problem. 
 
 The detailed analysis of NLO CF computation (see Refs. and references therein) shows that this instability comes from the high partonic center-of-mass energy ($\sqrt{\hat{s}}$) region of integration in the collinear factorisation formula, which for both considered processes can be written in a form:
 \begin{equation}
 \sigma(\sqrt{S})= \int\limits_{X_{\min}}^1 \frac{dX}{X} {\cal L}_{ij}(X,\mu_F) \hat{\sigma}_{ij}(X,\mu_R,\mu_F),  \label{eq:CF-formula}
 \end{equation}   
where $X=M^2/\hat{s}$, $X_{\min}=M^2/S$, for the case of hadroproduction, with $\sqrt{S}=\sqrt{s}$, the partonic luminosity ${\cal L}_{ij}$ is given by the convolution of PDFs of two colliding protons (see the Eq. (1.2) in Ref.~\cite{Lansberg:2021vie}) and partonic labels $i,j=\{g,q,\bar{q}\}$ In the photoproduction case, with $\sqrt{S}=\sqrt{s_{\gamma p}}$, we have just one proton PDF: ${\cal L}_{i\gamma }(X,\mu_F)=Xf_i(X,\mu_F)$. The $\hat{\sigma}_{ij}$ in Eq.~\ref{eq:CF-formula} is the CF coefficient function, which in the LO in $\alpha_s$ for the hadroproduction case is given by the partonic cross sections of the process:
\begin{equation}
g(q_1) + g(q_2) \to c\bar{c}\left[{}^1S_0^{[1]} \right](p), \label{subp:gg-cc}
\end{equation}
 with $\hat{s}=(q_1+q_2)^2$, $q_{1,2}^2=0$, so that $\hat{\sigma}_{gg}^{\text{(CF LO)}}\propto \delta(X-1)$. For the $J/\psi$ photoproduction process the LO contribution is given by:
\begin{equation}
g(q_1) + \gamma(q) \to c\bar{c}\left[{}^3S_1^{[1]} \right](p) + g, \label{subp:gag-ccg}
\end{equation}
 with $\hat{s}=(q_1+q)^2$, $q^2=q_1^2=0$, and $\hat{\sigma}_{g\gamma}^{\text{(CF LO)}}$ is a smooth function of $X$ in this case.
 
  The perturbative instability illustrated in Figs.~\ref{fig:sig-etac-CF} and \ref{fig:sig-jpsi-CF} arises due to the behaviour of the NLO CF coefficient function $\hat{\sigma}_{ij}^{\text{(CF NLO)}}(X,\mu_F,\mu_R)$ for $X\ll 1$, so it is natural to seek for a solution of this problem with perturbative resummation of the CF coefficient function in this region.

Such resummation is provided by the High-Energy Facotirsation (HEF) formalism of Refs.~\cite{Catani:1990xk,Catani:1990eg,Collins:1991ty,Catani:1994sq}, which resums the series of higher-order corrections to $\hat{\sigma}_{ij}$ at leading power in $X\ll 1$ which scale as $\alpha_s^n \ln^{n-1}(1/X)$, referred to as Leading Logarithmic Approximation (LLA). For the photoproduction case the resummation formula in the strict LLA($\ln(1/X)$) is derived in Ref.~\cite{Lansberg:2023kzf}:
\begin{eqnarray}
&& \hspace{-12mm} \frac{d\hat{\sigma}_{i\gamma}^{\text{(HEF)}}}{dz}(X,\mu_F,\mu_R)= \frac{1}{2M^2} \int\limits_0^\infty d{\bf k}_T^2 {\cal C}_{gi}(X,{\bf k}_T^2,\mu_F,\mu_R) \nonumber \\ &&\times \int\limits_{1/z}^\infty \frac{dy}{y} \frac{d{\cal H}}{dz}({\bf k}_T^2,y,z),
\end{eqnarray}
where we take into account the possibility of experimental cuts on the elasticity variable $z=(Pp)/(Pq)$, with $P$ being the proton momentum, and the resummation factor ${\cal C}_{gi}(X,{\bf k}_T^2,\mu_F,\mu_R)$ in the Doubly-Logarithmic Approximation (DLA) which resums terms $\propto \left[\alpha_s(\mu_R)\ln(1/X)\ln({\bf k}_T^2/\mu_F^2)\right]^n$, to stay consistent with $\mu_F$-evolution of PDFs, see the Sec. 2.3 of Ref.~\cite{Lansberg:2021vie} and references for more detailed discussion. The coefficient function $d{\cal H}/dz$ is derived in the Appendix A of Ref.~\cite{Lansberg:2023kzf} and is related to the following ``off-shell'' analog of the partonic subprocess (\ref{subp:gag-ccg}):
\begin{equation}
R_+(k)+\gamma(q)\to c\bar{c}\left[{}^3S_1^{[1]} \right](p) + g, \label{subp:Rga-ccg}
\end{equation}
with $k=q_1+k_T$ so that $k^2=-{\bf k}_T^2$ and $R_+$ denotes the {\it Reggeized gluon} which can be defined e.g. using the gauge-invariant EFT for Milti-Regge processes in QCD~\cite{Lipatov95}. Coefficient functions of subprocesses with one Reggeized gluon in the initial state, such as (\ref{subp:Rga-ccg}) are often referred to as {\it impact-factors} in literature.

The resummation formula for the $\eta_c$ hadroproduction case involves two resummation factors ${\cal C}_{gi}$ and is more cumbersome, so we will not reproduce it here, see the Eq. (2.6) in Ref.~\cite{Lansberg:2021vie}. Importantly, it involves the off-shell coefficient function which is given by the analog of partonic subprocess (\ref{subp:gg-cc}) with two Reggeised gluons in the initial state:
\begin{equation}
R_+(k_1) + R_-(k_2) \to c\bar{c}\left[{}^1S_0^{[1]} \right](p), \label{subp:RR-cc}
\end{equation}
with $k_{1,2}=q_{1,2}+k_{1,2T}$, $k_{1,2}^2=-{\bf k}_{1,2 T}^2$.

The HEF resummation outlined above is valid only for $X\ll 1$, so to compute the integral (\ref{eq:CF-formula}) we must combine it with the NLO CF approximation for $\hat{\sigma}_{ij}$ for $X\lesssim 1$. We do this using the smooth weight functions ($0<w_{ij}(X)<1$):
\begin{eqnarray}
&&\hspace{-10mm}\hat{\sigma}_{ij}(X) = w_{ij}(X) \hat{\sigma}_{ij}^{\text{(NLO CF)}}(X) \nonumber \\
&& + (1-w_{ij}(X)) \hat{\sigma}_{ij}^{\text{(HEF)}}(X),
\end{eqnarray}
 which are computed according to the Inverse Error Weighting (InEW) prescription of Ref.~\cite{Echevarria:2018qyi}, further developed in Refs~\cite{Lansberg:2021vie,Lansberg:2023kzf}.
 
\begin{figure}[t]
\begin{center}\includegraphics[width=0.4\textwidth]{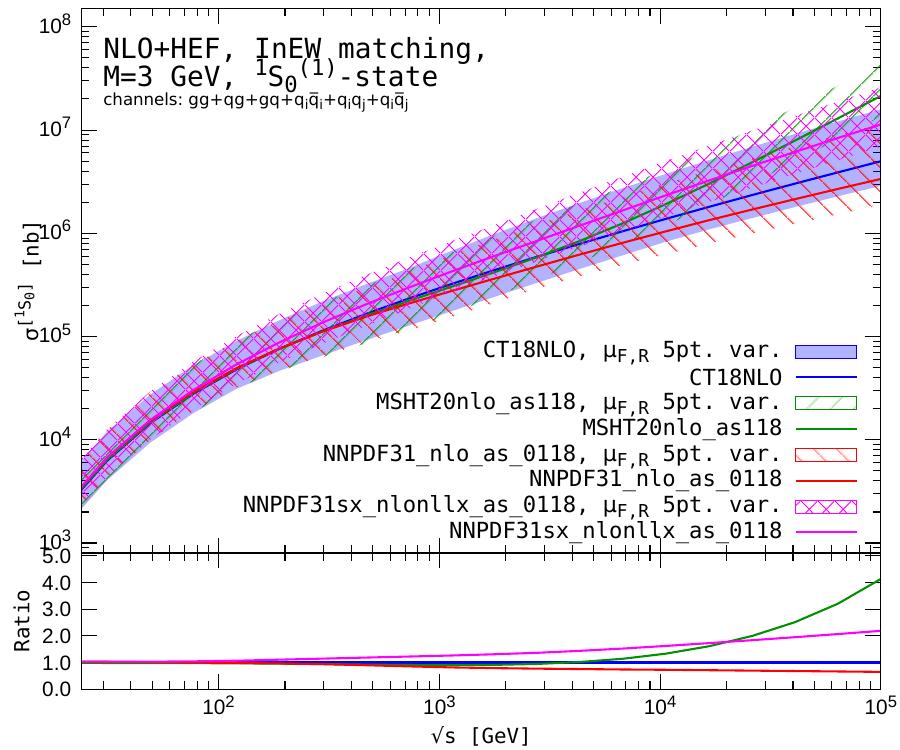}\end{center}
\caption{The same cross section as in the Fig.~\ref{fig:sig-etac-CF} computed via the matching of NLO computation in CF with the DLA HEF resummation in the approach of the Ref.~\cite{Lansberg:2021vie}. Different curves correspond to different PDF sets and bands correspond to the same 5 point scale variation as in the Fig.~\ref{fig:sig-etac-CF}.}\label{fig:sig-etac-match}
\end{figure}

\begin{figure}[t]
\begin{center}\includegraphics[width=0.4\textwidth]{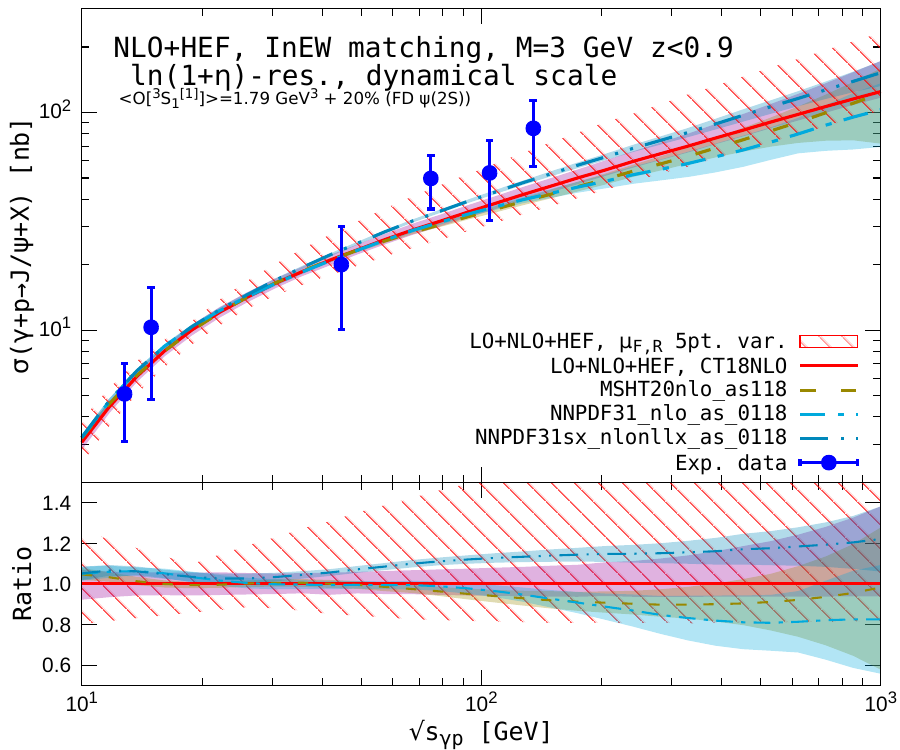}\end{center}
\caption{The same cross section as in the Fig.~\ref{fig:sig-jpsi-CF} computed via the matching of NLO computation in CF with the DLA HEF resummation in the approach of the Ref.~\cite{Lansberg:2023kzf} Different curves correspond to different PDF sets together with corresponding PDF uncertainties, shaded bands corresponds to the same 5 point scale variation as in the Fig.~\ref{fig:sig-jpsi-CF}.} \label{fig:sig-jpsi-match}
\end{figure}

 The results of such matched computation are shown in the Figs.~\ref{fig:sig-etac-match} and \ref{fig:sig-jpsi-match} for $\eta_c$ hadroproduction and $J/\psi$ photoproduction respectively. One can see, that the instability of scale-variation band at high energy is gone and the band is even reduced in comparison with the LO band shown in the Figs.~\ref{fig:sig-etac-CF} and \ref{fig:sig-jpsi-CF}. From the Fig.~\ref{fig:sig-jpsi-match} one can see, that PDF uncertainties for $\sqrt{s_{\gamma p}}<500$ GeV are clearly subdominant, showing that it is the improvement of high-energy behaviour of the CF coefficient function had stabilised the predictions. 
 
 These results look encouraging, however their residual scale uncertainty is still unacceptably large, calling for improvement of the computation beyond DLA. One of the key steps towards this goal is the computation of loop corrections to off-shell subprocesses such as (\ref{subp:Rga-ccg}) and (\ref{subp:RR-cc}). In the next section we report the first results of such computations involving NRQCD states, which we have performed.

\section{One-loop quarkonium impact factors}

In this section we present our computation of one loop impact-factors for the following processes:
\begin{eqnarray}
R_+(k) + \gamma(q) &\to& Q\bar{Q} \left[ {}^1S_0^{[8]} \right](p), \label{subp:Rga-1S08} \\
R_+(k) + g(q)&\to& Q\bar{Q} \left[ {}^1S_0^{[1]} \right](p), \label{subp:Rg-1S01}
\end{eqnarray}
where $q^2=0$, $k^2=-{\bf k}_T^2$, $q^+=k^-=0$, $q^-> 0$, $p^2=M^2=4m_c^2$ and light-cone components are defined as $k^\pm = k^0 \pm k^3$. The subprocess (\ref{subp:Rga-1S08}) is mostly of academic interest, since it contributes e.g. to the inclusive $J/\psi$ photoproduction at the ``exclusive'' kinematic threshold $z=1$ where no data exist. However it was very instructive for us to consider it, because of the smaller number of Feynman diagrams and master integrals contributing in comparison with the subprocess (\ref{subp:Rg-1S01}). The subprocess (\ref{subp:Rg-1S01}) is more physical and can be used to study $\eta_{c,b}$ production at forward rapidities at hadron colliders. 

However the most useful for the phenomenology at hadron colliders would be the computation of central production vertices, similar to the subprocess (\ref{subp:RR-cc}), which can be performed within the EFT formalism without computing new integrals, besides those which arise in the impact-factor computations. The computation reported in this proceedings serves as a stepping stone towards computation of central production vertices.   

\subsection{Outline of the computation}

\begin{figure}[t]
\includegraphics{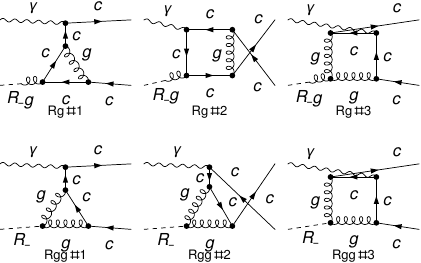}
\caption{Example Feynman diagrams with $Rg$ (top row) and $Rgg$ (bottom row) couplings, contributing to the subprocess (\ref{subp:Rga-1S08}) at one loop \label{fig:diags-1S08}}
\end{figure}

\begin{figure}[t]
\includegraphics{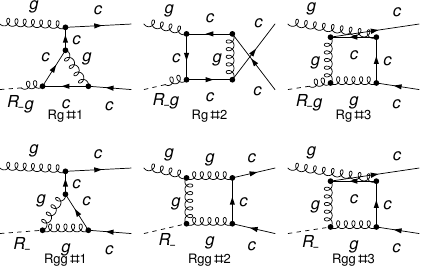}
\caption{Example Feynman diagrams with $Rg$ (top row) and $Rgg$ (bottom row) couplings, contributing to the subprocess (\ref{subp:Rg-1S01}) at one loop \label{fig:diags-1S01}}
\end{figure}

  The Feynman diagrams for both subprocesses had been generated using the custom made model file for \texttt{FeynArts}~\cite{FeynArts}, in which the $Rg$ ``mixing'' coupling (see e.g. Eq.~(13) in Ref.~\cite{Nefedov:2019mrg}) and $Rgg$ induced coupling (see e.g. Eq.~(14) in Ref.~\cite{Nefedov:2019mrg}) had been implemented. Example Feynman diagrams are shown in the Figs.~\ref{fig:diags-1S08} and~\ref{fig:diags-1S01}. Then, the NRQCD spin and colour projectors had been inserted, and momenta of heavy quarks had been put to $p_{c}=p_{\bar{c}}=p/2$ to project-out the $S$-wave. After taking the interference with corresponding LO impact-factor, obtained scalar quantity can be reduced down to one-loop master integrals using integration-by-parts (IBP) reduction, we use \texttt{FIRE}~\cite{Smirnov:2019qkx} package for this purpose. However, due to above-mentioned choice of momenta of heavy quarks, linearly-dependent quadratic denominators appear in some diagrams, e.g. in the $Rg$-coupling diagrams \#2 and \#3 and $Rgg$-coupling diagram \#3 in both Figs.~\ref{fig:diags-1S08} and \ref{fig:diags-1S01}. These linearly-dependent denominators have to be partial-fractioned into different master topologies before IBP reduction can be performed. This is the common procedure in one-loop computations involving quarkonia and we have implemented it into our \texttt{FeynCalc}~\cite{FeynCalc,Shtabovenko:2020gxv}-based code.    
  
 To regularise rapidity divergences in scalar one-loop integrals we tilt the direction-vectors of Wilson lines in the effective action from the light-cone:
 \begin{equation}
 n_\pm \to \tilde{n}_\pm = n_\pm + r n_\mp,
 \end{equation}
with $0<r\ll 1$ as was first proposed in Refs.~\cite{Hentschinski:2011tz,Chachamis:2012cc}, the same regularisation is also used in our Ref.~\cite{Nefedov:2019mrg}. Besides rapidity-divergent scalar one loop integrals which are listed in the latter paper the present computation also contains integrals mixing massive quadratic propagators with linear propagators, which also acquire nontrivial $r$-dependence. Fortunately, for such integrals massive propagators can be traded for massless ones, using the following algebraic identity:
\begin{eqnarray}
&&\hspace{-8mm}\frac{1}{((\tilde{n}_+ l) + k_+) (l^2-m^2)} = \frac{1}{((\tilde{n}_+ l) + k_+) (l+\kappa \tilde{n}_+)^2 } \nonumber \\
&&\hspace{-8mm} + \frac{2\kappa {\left[ (\tilde{n}_+ l)+\frac{m^2+\tilde{n}_+^2 \kappa^2}{2\kappa}\right]}}{{((\tilde{n}_+ l) + k_+)} (l+\kappa \tilde{n}_+)^2(l^2-m^2) },
\end{eqnarray}   
where we choose the parameter $\kappa$ in such a way that the linear denominator in the last term in the r.h.s. gets canceled, while in the first term we are left only with a linear and massless quadratic denominators. This identity can be applied recursively to remove all massive quadratic denominators from the integral containing the linear denominator. All additional terms generated by this procedure will be just usual one-loop integrals with quadratic massive or massless propagators. The known results from literature can be used for the latter integrals and we exploit the implementation of \texttt{PackageX}~\cite{Patel:2015tea} into \texttt{FeynHelpers}~\cite{Shtabovenko:2016whf} for this purpose. The new rapidity divergent scalar integrals which we have encountered during this computation are:
\begin{eqnarray}
&& \hspace{-14mm}B_{[-]}(-K,K-q)=\int \frac{d^D l}{[\tilde{l}^-](l-K)^2 (l+K-q)^2}, \\
&& \hspace{-14mm}C_{[-]}(0,-K,K-q)=\int \frac{d^D l}{[\tilde{l}^-] l^2 (l-K)^2 (l+K-q)^2}, \\
&& \hspace{-14mm}B_{[-]}(p,K)=\int \frac{d^D l}{[\tilde{l}^-](l+p)^2 (l+K)^2}, \\
&& \hspace{-14mm}C_{[-]}(p,K,k)=\int \frac{d^D l}{[\tilde{l}^-](l+p)^2 (l+K)^2 (l+k)^2},
\end{eqnarray}
where $K=[p - M^2 n_-/(2 q^-)]/2$ with $n_-^\mu=(1,0,0,1)^\mu$. These integrals have the same complexity as the integral $C_{[-]}$ with two scales, computed in the Ref.~\cite{Nefedov:2019mrg}. We will cover the computation of these integrals in more detail in a longer version of this paper.

\subsection{Results for quarkonium impact factors}

In this subsection we present results of the computation outlined above, which had been expanded in the limit $r\ll 1$ as well as in $\epsilon$. We present the real parts of interference of one-loop and LO impact-factors of subprocesses~(\ref{subp:Rga-1S08}) and (\ref{subp:Rg-1S01}), normalised by the corresponding LO impact-factors and with heavy-quark-mass renormalisation counterterm in the on-shell scheme added, which is customary for heavy quarkonium production studies. For subprocesses (\ref{subp:Rga-1S08}) and (\ref{subp:Rg-1S01}) respectively, these results can be written as follows:
\begin{eqnarray}
  &\hspace{-12mm} 2\Re\left[ \frac{H^{^1S_0^{[8]}}_{\text{1L x LO}}({\bf k}_T) + \text{(OS mass CT)} }{(\alpha_s/(2\pi))H^{^1S_0^{[8]}}_{\text{LO}}({\bf k}_T)} \right] = \left(\frac{\mu^2}{{\bf k}_T^2} \right)^{\epsilon} \frac{1}{\epsilon}\left[ - \frac{2n_F}{3} - \frac{3}{2N_c}  \right.  \label{eq:IF-1S08} \\
  &\hspace{-12mm} \left.  + N_c \left( \ln\frac{{\bf k}_T^2}{M^2} +\ln \frac{q_-^2}{{\bf k}_T^2 r} + \frac{19}{6} \right)  \right] + F_{{}^1S_0^{[8]}} ({\bf k}_T^2/M^2) + O(r,\epsilon), \nonumber \\
  &\hspace{-12mm} 2\Re\left[ \frac{H^{^1S_0^{[1]}}_{\text{1L x LO}}({\bf k}_T) + \text{(OS mass CT)} }{(\alpha_s/(2\pi))H^{^1S_0^{[1]}}_{\text{LO}}({\bf k}_T)} \right] = \left(\frac{\mu^2}{{\bf k}_T^2} \right)^{\epsilon} \left\{ -\frac{N_c}{\epsilon^2} + \frac{1}{\epsilon} \left[ - \frac{2n_F}{3} \right. \right. \label{eq:IF-1S01} \\
  &\hspace{-12mm} \left.\left. - \frac{3}{2N_c} +  N_c \left( \ln \frac{q_-^2}{{\bf k}_T^2 r} + \frac{25}{6} \right) \right]  \right\} + F_{{}^1S_0^{[1]}} ({\bf k}_T^2/M^2) + O(r,\epsilon). \nonumber
\end{eqnarray}
where $n_F$ is the number of flavours of light quarks. It is crucial, that the sole remaining dependence on $\ln r$ in Eqns.~(\ref{eq:IF-1S08}) and (\ref{eq:IF-1S01}) is proportional to the one-loop Regge trajectory of a gluon, as required by gluon Reggeisation, while terms $\sim\ln^2 r$ have canceled nontrivially between different diagrams. The remainder functions $F_{m}(\tau)$, with $m={}^1S_0^{[8]}$ or ${}^1S_0^{[1]}$, can be decomposed w.r.t. different colour structures:
\begin{equation}
\hspace{-5mm}F_{m}(\tau)=-\frac{10}{9}n_F +\Re[C_FF^{(C_F)}_{m}(\tau)+C_AF^{(C_A)}_{m}(\tau)], 
\end{equation}
The coefficients in front of $C_F$ are the same for both processes:
\begin{eqnarray}
   && \hspace{-5mm} F^{(C_F)}_{{}^1S_0^{[8]}}(\tau)=F^{(C_F)}_{{}^1S_0^{[1]}}(\tau)=\frac{ {\cal L}_2+{\cal L}_7(1-2 \tau)}{\tau +1} \nonumber \\
   &&\hspace{-5mm}+ \frac{1}{6 (\tau +1) (2 \tau +1)^2} \Bigl\{ 144 L_1 \tau ^2+144 L_1 \tau \nonumber
   \\
   &&\hspace{-5mm}  +36 L_1 -16 \pi ^2 \tau^3-72 \tau ^3+72 \tau ^3 \log (2) \nonumber \\
   &&\hspace{-5mm} -156 \tau ^2+12 \tau ^2 \log ^2(2 \tau
   +1)+168 \tau ^2 \log (2) \nonumber \\ 
   &&\hspace{-5mm}-24 \left(3 \tau ^2+5 \tau +2\right) \tau  \log
   (\tau +1) +12 \pi ^2 \tau \nonumber \\ 
   &&\hspace{-5mm}  -108 \tau +12 \tau  \log ^2(2 \tau +1)+3 \log
   ^2(2 \tau +1) \nonumber \\
   &&\hspace{-5mm}+132 \tau  \log (2) +18 (\tau +1) (2 \tau +1)^2 \log (\tau
   ) \nonumber \\ 
   &&\hspace{-5mm}+4 \pi ^2-24+36 \log (2)\Bigr\}.
\end{eqnarray}

The coefficient in front of $C_A$ for subprocess (\ref{subp:Rga-1S08}) is:
\begin{eqnarray}
&& \hspace{-10mm}      F^{(C_A)}_{{}^1S_0^{[8]}}(\tau)=\frac{1}{2 (\tau -1) (\tau +1)^3} \Bigl\{ (\tau +1)^2 \left(-4 {\cal L}_4 \left(\tau^2-1\right) \right. \nonumber \\
&& \hspace{-15mm}  \left. +{\cal L}_2 (\tau +1) (2 \tau +1)+{\cal L}_7 \tau  (2
   \tau -3)+{\cal L}_7\right) \nonumber \\
&& \hspace{-15mm}  +2 {\cal L}_6 (\tau  (\tau  ((\tau -4)
   \tau -6)-4)+1) \Bigr\} \nonumber \\
&& \hspace{-15mm} + \frac{1}{36 (\tau -1) (\tau +1)^3 (2
   \tau +1)} \Bigr\{ -216 L_1 \tau ^4-324 L_1 \tau ^3 \nonumber \\ 
&& \hspace{-15mm}  +108 L_1 \tau ^2+324
   L_1 \tau +108 L_1 +120 \pi ^2 \tau ^5 \nonumber \\ 
&& \hspace{-15mm}  +608 \tau ^5 -36 \tau ^5 \log ^2(\tau +1)+36 \tau ^5 \log ^2(2 \tau +1) \nonumber \\
&& \hspace{-15mm}  -36 \tau ^5 \log ^2(2) -72 \tau ^5 \log (2) \log (\tau +1) \nonumber \\  
&& \hspace{-15mm}  +216 \tau ^5 \log (\tau +1)+72 \tau ^5
   \log (2)+228 \pi ^2 \tau ^4 \nonumber \\
&& \hspace{-15mm} +1520 \tau ^4 -306 \tau ^4 \log ^2(2) +360 \tau ^4 \log (2)  \nonumber \\ 
&& \hspace{-15mm}  -306 \tau ^4 \log ^2(\tau +1)+144
   \tau ^4 \log ^2(2 \tau +1) \nonumber \\ %\end{eqnarray*}\begin{eqnarray}
&& \hspace{-15mm} +252 \tau ^4 \log (2)
   \log (\tau +1)+432 \tau ^4 \log (\tau +1) \nonumber \\ 
&& \hspace{-15mm} +84 \pi ^2 \tau ^3+608 \tau ^3 -360 \tau ^3 \log ^2(\tau +1) \nonumber \\ 
&& \hspace{-15mm}  -360 \tau ^3 \log ^2(2)+576 \tau ^3 \log (2) \log (\tau +1) \nonumber \\ 
&& \hspace{-15mm} +225 \tau ^3 \log ^2(2\tau +1) -1216 \tau^2 -108 \tau^2 \log ^2(2) \nonumber \\
&& \hspace{-15mm} +72 \tau ^3 \log (\tau +1)+72 \tau ^3 \log (2)-120 \pi ^2 \tau ^2\nonumber \\
&& \hspace{-15mm}  -108 \tau ^2 \log ^2(\tau +1) +171 \tau ^2 \log ^2(2 \tau +1) \nonumber \\
&&  \hspace{-15mm}  +504 \tau ^2 \log (2) \log (\tau +1)-360 \tau ^2 \log (2(\tau +1)) \nonumber \\
&& \hspace{-15mm}  -72 (\tau +1)^3 \left(2 \tau ^2-\tau -1\right)
   \log (\tau -1) \log (2/(\tau +1)) \nonumber \\
&& \hspace{-15mm}  +36 (2 \tau +1) \log (\tau )
   \left[-\tau ^4+\tau ^4 \log (8)-6 \tau ^2 \log (2) \right. \nonumber \\
&& \hspace{-15mm}  +\left(-\tau ^3+4 \tau
   ^2+6 \tau +4\right) \tau  \log (\tau +1) \nonumber \\
&& \hspace{-15mm}  \left. -8 \tau  \log (2)-\log (2 \tau
   +2)+1\right] +63 \tau  \log ^2(2 \tau +1)  \nonumber \\
&& \hspace{-15mm}  -18 \left(2 \tau ^5+17 \tau ^4+20 \tau ^3+6 \tau ^2-6 \tau
   -3\right) \log ^2(\tau ) \nonumber \\
&& \hspace{-15mm}  -84 \pi ^2 \tau -1216 \tau +108 \tau  \log
   ^2(\tau +1)\nonumber \\
&& \hspace{-15mm}  +108 \tau  \log ^2(2)+54 \log
   ^2(\tau +1)+9 \log ^2(2 \tau +1) \nonumber \\
&&  \hspace{-15mm}  +72 \tau  \log (2) \log (\tau +1)-288
   \tau  \log (\tau +1) \nonumber \\
&&  \hspace{-15mm}  -144 \tau  \log (2)-36 \log (2) \log (\tau +1) \nonumber \\
&&  \hspace{-15mm}  -72\log (\tau +1)-12 \pi ^2-304+54 \log ^2(2) \Bigr \}
\end{eqnarray}
while for the subprocess (\ref{subp:Rg-1S01}) it is:
\begin{eqnarray*}
   &&\hspace{-15mm} F^{(C_A)}_{{}^1S_0^{[1]}}(\tau)=\frac{1}{(\tau -1) (\tau +1)^3}\Bigl\{ 2 {\cal L}_1 \left(\tau ^2+\tau -2\right) (\tau +1)^3 \nonumber \\ 
   &&\hspace{-15mm}+\tau \Bigl[2 {\cal L}_5 \left(\tau  (\tau +1) \left(\tau
   ^2-2\right)+1\right) -{\cal L}_7 \left(\tau ^2+\tau
   -1\right) \nonumber \\ 
   &&\hspace{-15mm} -\left({\cal L}_2 (\tau +2) (\tau
   +1)^2\right) \nonumber \\
   &&\hspace{-15mm} +{\cal L}_6 (\tau  (\tau  (6-(\tau -4) \tau
   )+4)-1)\Bigr] \nonumber \\
   &&\hspace{-15mm} +2 {\cal L}_3 (\tau -1) (\tau +1)^3+2
   {\cal L}_5+{\cal L}_7 \Bigr\} \nonumber \\
   &&\hspace{-15mm}- \frac{1}{18 (\tau -1) (\tau +1)^3} \Bigl\{ 6 \pi ^2 \tau ^5-36 \tau ^5 \log (2) \log (\tau +1) \nonumber \\ \end{eqnarray*} \begin{eqnarray}  
   &&\hspace{-15mm}+36 \tau ^5 \log
   (\tau +1) \log (\tau +2)+63 \pi ^2 \tau ^4  -98 \tau ^4 \nonumber \\  
   &&\hspace{-15mm} -63 \tau ^4 \log^2(\tau +1)+9 \tau ^4 \log ^2(2 \tau +1) \nonumber \\
   &&\hspace{-15mm} -63 \tau ^4 \log ^2(2) +138 \pi ^2 \tau ^3 \nonumber \\  
   &&\hspace{-15mm} +54 \tau^4 \log (2) \log (\tau +1) -36 \tau ^4 \log (\tau +1) \nonumber \\ 
   &&\hspace{-15mm}+36 \tau ^4 \log
   (\tau +1) \log (\tau +2)+36 \tau ^4 \log (2) \nonumber \\
   &&\hspace{-15mm}-196 \tau^3-72 \tau ^3 \log ^2(\tau +1)+36 \tau ^3 \log ^2(2 \tau +1) \nonumber \\
   &&\hspace{-15mm} -72 \tau ^3\log ^2(2)+144 \tau ^3 \log (2) \log (\tau +1)\nonumber \\ 
   &&\hspace{-15mm} -36 \tau ^3 \log (\tau
   +1)-72 \tau ^3 \log (\tau +1) \log (\tau +2) \nonumber \\
   &&\hspace{-15mm} -36 \tau ^3 \log (2)+18 \pi
   ^2 \tau ^2 -18 \tau ^2 \log ^2(\tau +1) \nonumber \\ 
   &&\hspace{-15mm} +45 \tau ^2 \log ^2(2 \tau +1)-18
   \tau ^2 \log ^2(2)\nonumber \\ 
   &&\hspace{-15mm} +108 \tau ^2 \log (2) \log (\tau +1)  +36 \tau ^2 \log
   (\tau +1)\nonumber \\ 
   &&\hspace{-15mm} -72 \tau ^2 \log (\tau +1) \log (\tau +2)-36 \tau ^2 \log
   (2) \nonumber \\ 
   &&\hspace{-15mm} -18 \left(4 \tau ^4+5 \tau ^3+\tau ^2-3 \tau -1\right) \log ^2(\tau
   ) \nonumber \\ 
   &&\hspace{-15mm}+18 \log (\tau ) \Bigl[\tau ^5 \log (2)-\tau ^4 (\log (4)-2) \nonumber \\
   &&\hspace{-15mm} -2 \tau ^2 (1+\log (4)) -\tau ^3 \log (4) \nonumber \\
   &&\hspace{-15mm}-\left(\tau ^4-4 \tau ^3-6 \tau ^2-4 \tau
   +1\right) \tau  \log (\tau +1) \nonumber \\  
   &&\hspace{-15mm}-\tau  \log (8)-\log (4)\Bigr] \nonumber \\
   &&\hspace{-15mm} -120 \pi ^2
   \tau +196 \tau +36 \tau  \log ^2(\tau +1) \nonumber \\  
   &&\hspace{-15mm} +18 \tau  \log ^2(2 \tau +1)+36
   \tau  \log ^2(2)+9 \log ^2(\tau +1) \nonumber \\ 
   &&\hspace{-15mm} -36 \tau  \log (2) \log (\tau +1)+36
   \tau  \log (\tau +1) \nonumber \\
   &&\hspace{-15mm} +36 \tau  \log (\tau +1) \log (\tau +2)+36 \tau 
   \log (2) \nonumber \\ 
   &&\hspace{-15mm} -36 (\tau -1) (\tau +1)^3 \log (\tau -1) (\log (2)-\log (\tau
   +1)) \nonumber \\
   &&\hspace{-15mm} -18 \log (2) \log (\tau +1) +36 \log (\tau +1) \log (\tau +2) \nonumber \\ 
   &&\hspace{-15mm} -69 \pi^2+98+9 \log ^2(2) \Bigr \}.
\end{eqnarray}

In formulas above the following combinations of logarithms and dilogarithms appear:
\begin{eqnarray*}
  && \hspace{-15mm} L_1=\sqrt{\tau(1+\tau)}\ln\left[ 1+2\tau + 2\sqrt{\tau(1+\tau)} \right], \nonumber  \\
  && \hspace{-15mm} {\cal L}_1=\text{Li}_2\left(\frac{1}{\tau }+1\right), \nonumber \\
  && \hspace{-15mm} {\cal L}_2=\text{Li}_2\left(\frac{1}{-2 \tau -1}\right), \nonumber  \\
  && \hspace{-15mm}{\cal L}_3=\text{Li}_2\left(\frac{1}{\tau }\right)+\text{Li}_2\left(\frac{\tau
   -1}{\tau +1}\right)-\text{Li}_2\left(\frac{\tau +1}{2 \tau
   }\right) \nonumber \\  
  &&\hspace{-15mm} +\frac{\text{Li}_2\left(\frac{1}{4}\right)}{2}+\text{Li}_2(-2), \nonumber \\ 
  &&\hspace{-15mm} {\cal L}_4=\text{Li}_2\left(1+\frac{1}{\tau }\right)+\text{Li}_2\left(\frac{1}{\tau
   }\right)+\text{Li}_2\left(\frac{\tau -1}{\tau
   +1}\right) \nonumber \\ 
  &&\hspace{-15mm} -\text{Li}_2\left(\frac{\tau +1}{2 \tau
   }\right)+\frac{\text{Li}_2\left(\frac{1}{4}\right)}{2}+\text{Li}_2(-2), \nonumber  \\ \end{eqnarray*} \begin{eqnarray*}
  &&\hspace{-15mm} {\cal L}_5 = \text{Li}_2\left(-\frac{1}{\tau +1}\right)-\text{Li}_2(\tau +2)+\frac{1}{2}
   \text{Li}_2\left(\frac{2 \tau +1}{2 \tau +2}\right), \nonumber  \\ 
  &&\hspace{-15mm} {\cal L}_6 =-\text{Li}_2\left(-\frac{2 \tau +1}{\tau
   ^2}\right)+\text{Li}_2\left(-\frac{-2 \tau ^2+\tau +1}{2 \tau
   ^2}\right) \nonumber \\ 
  &&\hspace{-15mm} +\text{Li}_2\left(\frac{1}{2}-\frac{\tau
   }{2}\right)+\text{Li}_2\left(-\frac{1}{\tau
   }\right) \nonumber \\ 
  &&\hspace{-15mm} -\text{Li}_2\left(\frac{\tau -1}{2 \tau
   }\right)-\text{Li}_2(-\tau )+\text{Li}_2\left(\frac{1-\tau }{\tau
   +1}\right), \nonumber  \\
  &&\hspace{-15mm} {\cal L}_7 =\text{Li}_2(-2 \tau -1)-\text{Li}_2\left(\frac{2 \sqrt{\tau }}{\sqrt{\tau
   }-\sqrt{\tau +1}}\right) \nonumber \\
  &&\hspace{-15mm} -\text{Li}_2\left(\frac{2 \sqrt{\tau
   }}{\sqrt{\tau }+\sqrt{\tau +1}}\right).
\end{eqnarray*}
The number of dilogarithms with different arguments can clearly be reduced using known dilogarithm identities to reveal further structure of impact-factor expressions.

\section{Conclusions and Outlook}
In this contribution we have discussed the perturbative instability of $p_T$-integrated cross sections of production of heavy quarkonia at NLO of CF and its resolution through the matching with DLA resummation in the HEF formalism. We also describe our progress towards going beyond DLA, namely the first computation of one-loop corrections to impact factors involving NRQCD states of the $Q\bar{Q}$ pair: $Q\bar{Q}\left[{}^1S_0^{[8]} \right]$ and $Q\bar{Q}\left[{}^1S_0^{[1]} \right]$. The expected structure of rapidity, ultraviolet and infrared divergences had been found, which is a strong cross-check of the computation. In future the real-emission contribution will be also computed to obtain the infrared finite NLO correction to the impact-factors. 

{\bf Acknowledgments:} This work is supported by the Marie Sk{\l}odowska-Curie action ``RadCor4HEF'' under grant agreement No.~101065263.

\bibliography{mybibfile}

\end{document}